\documentclass[letterpaper,titlepage,12pt]{article}
\usepackage[utf8]{inputenc} 
\usepackage{textcomp} 
\usepackage{graphicx}  
\usepackage{flafter}  

\usepackage{amsmath,amssymb}  
\usepackage{bm}  
\usepackage{wasysym}

\usepackage{memhfixc}  
\usepackage{pdfsync}  
\begin{document}
\title{On the Lamb vector divergence, evolution of pressure fields and Navier-Stokes regularity}
\author{Jussi Lindgren, University of Helsinki, jussi.i.lindgren(at)helsinki.fi}
\maketitle
\begin{abstract}
This paper analyzes the Lamb vector divergence, also called the hydrodynamic charge density, and its implications to the Navier-Stokes system. It is shown that the pressure field can be always chosen in a way that ensures regularity of the Navier-Stokes system. The abstract pressure field that ensures regularity is defined through two partial differential equations, being of the elliptic kind. The pressure field defined such a way can be interpreted as a control potential field that keeps the system regular. The controlling pressure field depends only on the velocity field of the fluid and its derivatives, so that the result is applicable in any general setting where the initial data is divergence free, smooth and square-integrable.
\end{abstract}
\section{Introduction}
The Navier-Stokes equations are supposed to model any kind of fluid flow, including turbulent ones. The essential problem with the equations is the nonlinearity of the convective term. The pressure gradient represents transport and the diffusive term represents loss of energy in the system. In other words, to understand the equations, one has to understand the nonlinear convection term. The Navier-Stokes equations can be written in vector form as follows
\begin{equation}
\frac{\partial \vec{u}}{\partial t}+\vec{u}\cdot \nabla \vec{u}=\nu\Delta \vec{u}-\nabla {p}
\end{equation}
where $\vec{u}(x,y,z,t)\in \mathbb{R}^3$ is the velocity field of the fluid and $p(x,y,z,t)\in \mathbb{R}$ is the scalar pressure field. The diffusive term depends on a constant $\nu >0$.
The convective term is $\vec{u}\cdot \nabla \vec{u}$. This term can be decomposed according to the vector calculus identities
\begin{equation}
\vec{u} \cdot \nabla \vec{u}=\vec{\omega} \times \vec{u}+\nabla (\frac{1}{2}\vec{u}\cdot \vec{u})
\end{equation}
Subsituting this into the Navier-Stokes equations, one has
\begin{equation}
\frac{\partial \vec{u}}{\partial t}=-\vec{\omega}\times \vec{u}-\nabla (p+\frac{1}{2}\vec{u}\cdot \vec{u})+\nu\Delta \vec{u}
\end{equation}
where $\vec{\omega}=\nabla \times \vec{u}$ is the vorticity of the fluid flow. This formulation of the Navier-Stokes equations is known as the Lamb formulation \cite{lamb}.

In addition we assume that the fluid flow is incompressible, that is 
\begin{equation}
\nabla \cdot \vec{u}=0
\end{equation}
The system is assumed to have some smooth and square-integrable initial data $\vec{u}^0=\vec{u}(x,y,z,0)$ over the whole space $\mathbb{R}^3$.

\section{The regularity of Navier-Stokes equations}
When we say that the solutions are regular, we mean that two conditions are fulfilled, see \cite{clay}.
\begin{equation}
\vec{u},p\in C^{\infty}(\mathbb{R}^3\times [0,\infty))
\end{equation}
and
\begin{equation}
\int_{\mathbb{R}^3}\mid \vec{u}\mid ^2 dx <C
\end{equation}
for all $t\geq 0$. It is a well known fact that the solutions are regular if the enstrophy of the system stays bounded, see \cite{lulu}. This means that we will require
\begin{equation}
\int_{\mathbb{R}^3}\mid \vec{\omega}\mid ^2 dx <c
\end{equation}
for all $t\geq 0$. This implies that it is sufficient that the vorticity does not blow up. 
\section{The Lamb vector and its divergence}
The Lamb vector $L$ is the nonlinear term in the Lamb formulation of the Navier-Stokes equations
\begin{equation}
L=\vec{\omega}\times \vec{u}
\end{equation}
By applying the divergence operator to the Navier-Stokes equations, one has for the divergence of the Lamb vector
\begin{equation}
\nabla \cdot (\vec{\omega}\times \vec{u})=-\Delta (p+\frac{1}{2}\vec{u}\cdot \vec{u})
\end{equation}
Where the Laplacian is the normal, or scalar Laplacian. Marmanis and Shridar have called the divergence of the Lamb vector as hydrodynamic charge density, see \cite{marma}, \cite{shridar} in analogy to the electromagnetic theory.
\subsection{The divergence theorem and Lamb vector divergence}
The Lamb vector divergence can be represented as a sum of two parts
\begin{equation}
\nabla \cdot L=\vec{u}\cdot \nabla \times \vec{\omega}-\vec{\omega}\cdot \vec{\omega}
\end{equation}
,see an excellent paper on Lamb vector divergence by Hamman et al \cite{kirby}. The first term, $\vec{u}\cdot \nabla \times \vec{\omega}$ is called the flexion product whereas the second term is related to enstrophy.  From the vector calculus, we have the divergence theorem:
\begin{equation}
\iiint_V (\nabla \cdot \vec{F})dV=\oiint _S\vec{F} \cdot \vec{S}
\end{equation}
where $V$ is a compact subset of $\mathbb{R}^3$ and $\vec{F}$ is some continuously differentiable vector field in $V$. The vector $\vec{S}$ is the normal vector to a smooth surface embedding $V$. 
Moreover, from the vector calculus identities we have
\begin{equation}
\iiint_V(\nabla \cdot L)dV=\iiint_V(\vec{u}\cdot \nabla \times \vec{\omega}-\vec{\omega}\cdot \vec{\omega})dV=\oiint _S(\vec{\omega}\times \vec{u}) \cdot \vec{S}
\end{equation}
Now let us substitute the explicit Lamb divergence on the left side of the equation
\begin{equation}
\iiint_V(-\Delta (p+\frac{1}{2}\vec{u}\cdot \vec{u}))dV=\iiint_V(\vec{u}\cdot \nabla \times \vec{\omega}-\vec{\omega}\cdot \vec{\omega})dV
\end{equation}
We note immediately that we can obtain the local enstrophy from the equation 
\begin{equation}
\iiint_V\vec{\omega}\cdot \vec{\omega}dV=\iiint_V(\vec{u}\cdot \nabla \times \vec{\omega})dV+\iiint_V(\Delta (p+\frac{1}{2}\vec{u}\cdot \vec{u}))dV
\end{equation}
Putting the right side under the same integral
\begin{equation}
\iiint_V\vec{\omega}\cdot \vec{\omega}dV=\iiint_V(\vec{u}\cdot \nabla \times \vec{\omega}+\Delta (p+\frac{1}{2}\vec{u}\cdot \vec{u}))dV
\end{equation}
\section{Pressure as stabilizing force ensuring regularity}
Now, in order to prevent a blow-up, a sufficient condition is to have the enstrophy bounded in $\mathbb{R}^3$, so we want the left side of the equation to be bounded from above. We will proceed in two steps. First, let us suppose now that the pressure field $p$ is composed of two parts as follows
\begin{equation}
p(x,y,z,t)=q(x,y,z,t)+r(x,y,z,t)
\end{equation}
Substituting this into the integral equation yields
\begin{equation}
\iiint_{\mathbb{R}^3}\vec{\omega}\cdot \vec{\omega}dV=\iiint_{\mathbb{R}^3}(\vec{u}\cdot \nabla \times \vec{\omega}+\Delta (q+r+\frac{1}{2}\vec{u}\cdot \vec{u}))dV
\end{equation}
Taking the Laplacian term by term
\begin{equation}
\iiint_{\mathbb{R}^3}\vec{\omega}\cdot \vec{\omega}dV=\iiint_{\mathbb{R}^3}(\vec{u}\cdot \nabla \times \vec{\omega}+\Delta q+\Delta r+\Delta (\frac{1}{2}\vec{u}\cdot \vec{u}))dV
\end{equation}
We want to ensure that the first component of the pressure field partly balances the stress from the convection terms by setting
\begin{equation}
\Delta q=-\vec{u}\cdot \nabla \times \vec{\omega}-\Delta (\frac{1}{2}\vec{u}\cdot \vec{u})
\end{equation}
This pressure field $q$ is well defined through the Poisson-like equation. What is left is
\begin{equation}
\iiint_{\mathbb{R}^3}\vec{\omega}\cdot \vec{\omega}dV=\iiint_{\mathbb{R}^3}\Delta r dV
\end{equation}
Differentiating with respect to time we have,
\begin{equation}
\frac{d}{d t}\iiint_{\mathbb{R}^3}\vec{\omega}\cdot \vec{\omega}dV=\frac{d}{d t}\iiint_{\mathbb{R}^3} \Delta r dV
\end{equation}
This is the control equation for the evolution of enstrophy. We now require that this satisfies 
\begin{equation}
\frac{d}{d t}\iiint_{\mathbb{R}^3}\Delta r dV\le 0
\end{equation}
There is at least one expression that makes this possible and that depends only on the velocity field, namely the time evolution of the total kinetic energy $K(t)$
\begin{equation}
\frac{dK(t)}{dt}=\frac{d}{dt}\frac{1}{2}\int_{\mathbb{R}^3} \mid \vec{u}\mid ^2dV \le 0
\end{equation}
So let us choose 
\begin{equation}
\Delta r=\frac{1}{2}\mid \vec{u}\mid ^2
\end{equation}
Now we have
\begin{equation}
\frac{d}{d t}\iiint_{\mathbb{R}^3}\vec{\omega}\cdot \vec{\omega}dV=\frac{d}{dt}\frac{1}{2}\int_{\mathbb{R}^3}\mid \vec{u}\mid ^2dV \le 0
\end{equation}
That is, the evolution of enstrophy is equal to the evolution of total kinetic energy. To take stock, we have a pressure field $p$ such that
\begin{equation}
p(x,y,z,t)=q(x,y,z,t)+r(x,y,z,t)
\end{equation}
such that the pressure components satisfy
\begin{equation}
\Delta q=-\vec{u}\cdot \nabla \times \vec{\omega}-\Delta (\frac{1}{2}\vec{u}\cdot \vec{u})
\end{equation}
and
\begin{equation}
\Delta r=\frac{1}{2}\mid \vec{u}\mid ^2
\end{equation}
Putting them together
\begin{equation}
\Delta p=\Delta(q+r)=-\vec{u}\cdot \nabla \times \vec{\omega}-\Delta (\frac{1}{2}\vec{u}\cdot \vec{u})+\frac{1}{2}\mid \vec{u}\mid ^2
\end{equation}
This Poisson equation for the pressure ensures that the time evolution of enstrophy satisfies
\begin{equation}
\frac{d}{d t}\iiint_{\mathbb{R}^3}\vec{\omega}\cdot \vec{\omega}dV\le 0
\end{equation}
which in turn implies regularity for the Navier-Stokes system with smooth, divergence-free and square-integrable initial data.
\section{Consequences on the energy dissipation rate }
It is a well known fact, see \cite{lulu} that the following equality holds
\begin{equation}
\frac{dK}{dt}=-\nu E
\end{equation}
where $K(t)=\frac{1}{2}\int_{\mathbb{R}^3}\mid \vec{u}\mid ^2dV$ is the total kinetic energy of the system and $E(t)=\int_{\mathbb{R}^3}\mid \vec{\omega} \mid ^2dV$ is the total enstrophy of the system.
Let us differentiate the energy evolution equation with respect to time
\begin{equation}
\frac{d^2K(t)}{dt^2}=-\nu\frac{ dE(t)}{dt}
\end{equation}
and by noting that $\frac{dK(t)}{dt}=\frac{dE(t)}{dt}$ we will have the result
\begin{equation}
\frac{d^2K(t)}{dt^2}=-\nu\frac{ dK(t)}{dt}
\end{equation}
which has the general solution
\begin{equation}
K(t)=e^{rt}
\end{equation}
where $r=0$ or $r=-\nu$. The case where $r=0$ refers to the specific case where the initial data has zero vorticity. To exclude that trivial case, we assume $r=-\nu$ and thus have
\begin{equation}
K(t)=e^{-\nu t}
\end{equation}
so that the total kinetic energy of the system and thus vorticity decay exponentially to zero as $t \longrightarrow \infty$.

\end{document}